\documentclass[prb,aps,twocolumn,nofootinbib,eqsecnum,superscriptaddress]{revtex4-2}
\usepackage[colorlinks=true,linkcolor=blue,citecolor=blue,urlcolor=blue]{hyperref}
\usepackage{inputenc,amsfonts,amsmath,amsthm,amssymb,amsmath,amscd,graphicx,mathrsfs,braket,xcolor,dsfont,physics,comment}

\def\mO{\mathcal{O}}

\renewcommand\vec{\boldsymbol}

\usepackage{marginnote}

\usepackage{ulem}

\begin{document}
    \title{Stability of sub-dimensional localization to electronic interactions}

\author{Nisarga Paul}
\email{npaul@mit.edu}
\affiliation{Department of Physics, Massachusetts Institute of Technology, Cambridge, MA, USA}
\author{Philip J.D. Crowley}
\affiliation{Department of Physics, Harvard University, Cambridge, MA, USA}

\begin{abstract}
    Sub-dimensional localization, also known as directional localization, arises when 2d electrons are subject to a periodic potential and incommensurate magnetic field which cause them to become exponentially localized along one crystal axis, while remaining extended in the orthogonal direction. 
    We establish that sub-dimensional localization is robust to the presence of electronic interactions, which are known to generically destabilize single-particle localization.
    Consequently we predict that sub-dimensional localization may occur in moir\'e materials, where it is manifest as a total absence of conductivity in the localized (transverse) direction, with finite conductivity in the extended (longitudinal) direction. 
    We find the electrons form longitudinal charge density waves
    which are incommensurate with the moir\'e potential, allowing them to slide and hence conduct in the longitudinal direction. The conductivity remains prohibited in the transverse direction due an emergent charge dipole moment conservation law.
\end{abstract}
\maketitle

\section{Introduction}

Anderson showed that random disorder can lead to the localization of single electron orbitals, and hence the absence of dc conductivity~\cite{Anderson1958Mar}. 
Whether this phenomenology survives in the presence of weak electron-electron interactions has been the topic of extensive study~\cite{Basko2006May,Oganesyan2007Apr,Huse2014Nov,Nandkishore2015Mar,Abanin2019May}, with analyses indicating that dc transport is restored, except in one-dimensional system above a critical disorder strength~\cite{imbrie2016many,de2017stability,dumitrescu2019kosterlitz,sels2022bath,de2024absence}.

Single particle localization may also be induced without disorder, such as by an electric field~\cite{wannier1962dynamics}, an incommensurate periodic potential~\cite{AAH,Azbel:1979aa,Sokoloff1980Dec,Bourgain2007Sep,Crowley2018Apr}, or by a uniform magnetic field incommensurate with the underlying lattice~\cite{Hofstadter:1976aa,Harper:1955kl,Barelli1999Dec}. 
Similar incommensuration is possible in a moir\'e material, when the magnetic field is incommensurate with the moire potential, or quantitatively when the ratio of the magnetic length $\ell = \sqrt{\hbar/eB}$ and moire period $a_\mathrm{M}$ is irrational. When the moir\'e material is anisotropically strained, this incommensuration leads to \textit{sub-dimensional localization}, also known as \textit{directional localization}, specifically that the dc conductivity is non-zero only along certain crystallographic axes, and zero in the perpendicular direction.
%
Significantly, the directional localization is an interference effect, and \textit{not} a geometric effect of the strain, as indicated by the localization direction changing discretely and periodically with the inverse field strength $1/B$~\cite{Paul2024Jun}.

In this paper we show that remarkably this directional localization survives in the presence of weak electron-electron interactions. 
Specifically, we consider interacting electrons in a Landau level subject to a periodic potential and external magnetic field. 
We analyze the conductivity, and show for generic filling fractions, the directional localization is stable to the introduction of weak interactions.
This is achieved via a renormalization group (RG) analysis of a low energy effective model, consisting of coupled wires. For analytical control, our results are restricted to the regime of strong field and weak interactions (ie. Coulomb interaction $\ll$ moir\'e potential $\ll$ cyclotron frequency) though we anticipate, as in the single particle case, that the directional localization persists beyond this regime.

Our results are surprising, not only due to the generic fragility of localization due to interactions, but due to the na\"ive expectation that for weak interactions under RG a smectic metal should flow to a 2d Fermi liquid \cite{Emery2000Sep,Giamarchi2004}, and hence we should expect a non-zero conductivity in all directions. Instead, we find charge-density-wave (CDW) order, with conductivity in a single direction whose current is carried by a translational Goldstone mode. 

Our results show that the extensively studied physics of sub-dimensional particles~\cite{Nandkishore2019Mar,Pretko2020Feb} is accessible in moir\'e materials.
Specifically, we find electronic interactions are unable to restore charge transport in the transverse direction due an emergent charge dipole moment conservation law which constrains the mobility of the electrons. 
Such conservation laws are known to lead to anomalous dynamics including glassy behaviour~\cite{Prem2017Apr}, ergodic breaking~\cite{schulz2019stark,Sala2020Feb,Khemani2020May,morong2021observation}, and fracton hydrodynamics~\cite{Han2023Apr,Gromov2020Jul,Zerba2024Oct}.

%

This paper proceeds as follows. In \textsection \ref{sec:review} we review directional localization. In \textsection \ref{sec:eft} we translate this into a problem of interacting coupled wires, and subsequently establish the survival of the zero-transverse (\textsection \ref{sec:transverse}) and non-zero-longitudinal conductivity (\textsection \ref{sec:parallel}). We support these claims with a numerical analysis of the phase diagram \textsection \ref{sec:phases}, before concluding \textsection \ref{sec:disc}.

\section{Directional localization}
\label{sec:review}

We first review directional localization  following Ref. \cite{Paul2024Jun}, offering two complementary viewpoints on the underlying physics from the strong field (\textsection\ref{sec:strongfieldpicture}) and weak field (\textsection \ref{sec:levelsetpicture}) limits. Our starting point is a 2D electron gas (2DEG) in an out-of-plane magnetic field $B$ subject to a moir\'e potential $V(\vec r)$ with characteristic scale $V_0$
\begin{equation}
    H = H_0(\vec p- e \vec A) + V(\vec r)
    \label{eq:H0}
\end{equation}
where $\hbar = 1$, and $\vec\nabla \times \vec A = B\hat z$. 

Though not required for the results obtained, for simplicity we restrict to the case of quadratic dispersion $H_0(\vec p) = p^2/2m$ and a scalar potential with Fourier expansion
\begin{equation}
    V(\vec r) = \sum_i V_i e^{i\vec Q_i \cdot \vec r}
\end{equation}
as is generic in moir\'e systems, and we assume the potential to be rotationally symmetric with $C_n$ symmetry $n=3,4,$ or $6$.

\subsection{Strong field picture}
\label{sec:strongfieldpicture}

In the strong field picture, the cyclotron energy is the largest energy scale. Then the diagonal states of $H_0$ can be understood as the conventional Landau levels (with spacing $\omega_c = eB/m$) broadened into bands of width $V_i \ll \omega_c$ by the moir\'e potential. In this regime, directional localization follows from first-order perturbation theory in $V_i/\omega_c$. \par 
Projecting the full Hamiltonian~\eqref{eq:H0} into the $n$th Landau level one obtains
\begin{equation}
    \tilde V^{(n)} = \sum_i \tilde V_i e^{i \vec Q_i \cdot \tilde{\vec r}}
\end{equation}
where we have neglected the constant energy offset $\omega_c (n + \tfrac12 )$ from the bare Landau Level, and furthermore have defined the position operator projected into the $n$th Landau level $\tilde{\vec r} = (\tilde x,\tilde y)$ which satisfies
\begin{equation}
    [\tilde x,\tilde y] = i\ell^2
\end{equation}
and the projected Fourier coefficients are given by 
\begin{equation}\label{eq:tildeV}
    \tilde V_i = V_i L_n(Q_i^2\ell^2/2)e^{-Q_i^2\ell^2/4}
\end{equation}
where $L_n$ is the $n$th Laguerre polynomial. Importantly, the projected potential is strictly weaker $|\tilde V_i| \leq |V_i|$. Furthermore, the $\tilde V_i$ may be tuned to zero for special values of $Q_i \ell$ (corresponding to the $n$ zeroes of the $n$th Laguerre polynomial).

\par 

Directional localization is possible only in the absence of any rotational symmetry $C_n$ with $n \geq 2$. To break this symmetry of the moire potential, we consider the behavior of the system under strain. To describe the strain we simply to apply a linear transformation sending $\vec Q_i \mapsto \vec M \vec Q_i$. Both uniaxial strain [$\vec M = \text{diag}(\alpha,1/\alpha)$] or shear strain [$\vec M = \begin{pmatrix}
        1 & \beta \\ 0 & 1
\end{pmatrix}$] can be considered. 
This breaks the $C_n$ symmetry.

While directional localization is generic, it is most simply illustrated for the case of uniaxial strain applied along the crystallographic axis of a $C_4$ lattice. For simplicity, we further adopt the standard lowest harmonic (or first shell) approximation 
\begin{equation} \label{eq:Vr}
    V(\vec r) = V_x \cos(Q_x x) + V_y \cos(Q_y y)
\end{equation}
thus, after applying strain we have two reciprocal lattice vectors $\vec{Q}_x = Q_x \hat{\vec{x}}$, $\vec{Q}_y = Q_y \hat{\vec{y}}$. The effective Hamiltonian is then 
\begin{equation}\label{eq:AAH1}
    \tilde V^{(n)}(\tilde{\vec r}) = \tilde V_x\cos(Q_x \tilde x) +\tilde V_y \cos(Q_y \tilde y).
\end{equation}
Generically, the strain causes the projected potential to be stronger in one direction $|\tilde V_x| \neq |\tilde V_y|$. One finds that the eigenstates are consequently extended in the direction of the weaker potential and exponentially localized in the perpendicular direction; this is directional localization. In the simple $C_4$ example considered here, the localization can be proven via a direct mapping to the canonical Aubry-Andr\'e-Harper model~\cite{Paul2024Jun}.

The directional localization holds for generic values of the flux, specifically when the flux per unit cell is irrational\footnote{technically, Diophantine \cite{Jitomirskaya1999} This is only a slight restriction on irrationals, however, and Diophantine numbers are still dense in the real line.}. When the flux per unit cell is rational, there exists a finite magnetic unit cell, and the eigenstates are always extended in all directions. \par 

We will continually come back to this simple example, although many generalizations (to $C_6$ superlattices, shear strain, higher harmonics, matrix moir\'e potentials, etc.) exist. We also define some terminology here. It follows from Eq. \eqref{eq:tildeV} that there exist conditions under which either $\tilde V_x = 0$ or $\tilde V_y = 0$. We refer to the discrete set of parameters for which $\tilde V_x = 0$ or $\tilde V_y = 0$ as the \textit{maximally anisotropic points}.

\subsection{Level set picture}
\label{sec:levelsetpicture}

\begin{figure}
    \centering
\includegraphics[width=0.8\linewidth]{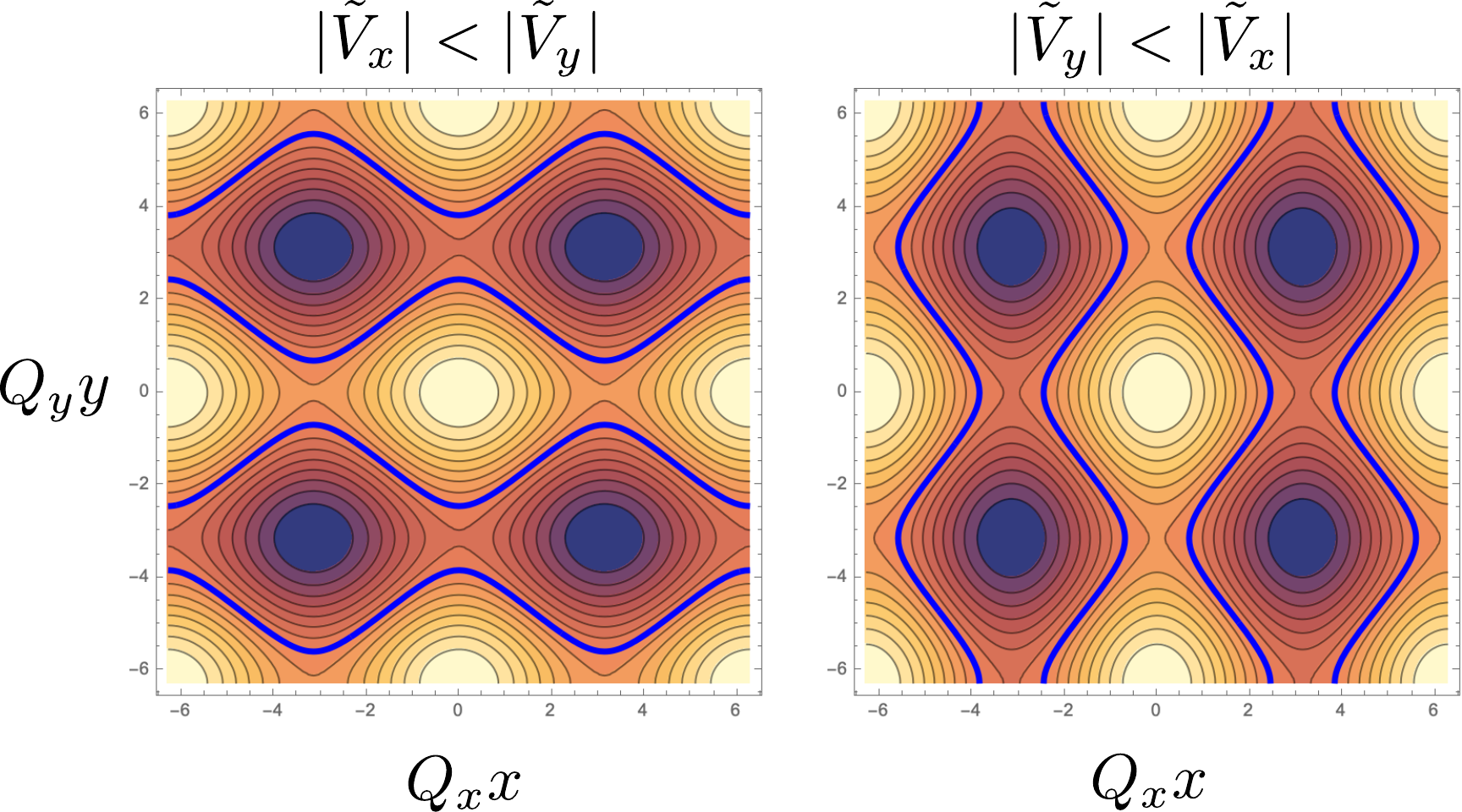}
    \caption{Level sets of the functions $\tilde V_x \cos (Q_x x) + \tilde V_y\cos (Q_y y)$ for $\tilde V_x = 1, \tilde V_y = 1.3$ [left panel] and vice versa [right panel] with some open level sets highlighted.}
    \label{fig:contours}
\end{figure}

A complementary, intuitive picture of directional localization comes from plotting the level sets of Eq. \eqref{eq:AAH1}. This is shown in Fig. \ref{fig:contours}. In the absence of strain, there are no open level sets. When $|\tilde V_y|< |\tilde V_x|$, however, there are open level sets along $y$, and vice versa for $|\tilde V_x|< |\tilde V_y|$. If we adopt a semiclassical picture, and treat $\tilde x$ and $\tilde y$ as (approximately) commuting (classical) variables, and combine this with semiclassical reasoning, we conclude that electrons undergo cyclotron motion tightly localized to the these level sets and guiding center drift along the level sets \cite{Beenakker1989Apr}. Then the presence of open level sets, absent without strain and generically present with strain, clearly indicates that some states extend along one direction while remaining localized along the transverse direction. 

This picture applies most straightforwardly in the limit $\ell \ll 1/Q_x, 1/Q_x$, where $[\cos(Q_x\tilde x),\cos(Q_y\tilde y)]\approx 0$, where $\tilde x$ and $\tilde y$ may indeed be treated approximately commuting variables, however this reasoning is known to apply in the quantum case too~\cite{wilkinson1984critical,wilkinson1987exact,han1994critical,yeo2022non}.

\par 

The level set picture provides a simple diagnostic for determining the axis along which states are extended for any set of microscopic conditions. This is helpful in practice, for instance when the principal axes of the strain are not controllable. \par 

Moreover, the level set picture illustrates that directional localization can be quite robust to single-particle perturbations. Sufficiently small perturbations (as might arise from strain disorder) will not alter the direction of the open level sets, and hence will not affect the directional localization. This is the result of a ``topological" robustness of level sets of potential landscapes \cite{Novikov1996May}. \par

Directional localization has a direct signature in the conductivity. Specifically, the single particle (i.e. Kubo) conductivity is zero in the transverse (localized) direction, and finite in the longitudinal (extended) direction~\cite{Paul2024Jun}. In the remaining sections, we show that this conclusion is robust to the inclusion of Coulomb interactions.

\section{Coupled wires effective theory}
\label{sec:eft}

The level set picture of the previous section motivates us to use a coupled wire model to studying the fate of directional localization in the presence of interactions: If the band is partially filled so that the open level sets occur at the Fermi level, then these are chiral quantum wires which couple via Coulomb interactions. In this section, we employ this model to study the fate of directional localization in the presence of interactions. 

In detail, we study interactions in the vicinity of maximally anisotropic points, at interaction strengths weak compared to the projected modulation strengths $\tilde V_i$, and show that the directional localization is stable.

\par 
All of the essential physics is present in the simplest case of the square superlattice case with uniaxial heterostrain along a principal axis. Therefore, for simplicity, we treat only this case in detail. As before, the bar models is given by~\eqref{eq:H0} with a quadratic dispersion and a square lattice moir\'e potential~\eqref{eq:Vr} with $Q_x\neq Q_y$ due to strain. We work in the strong field regime, where 
\begin{equation}\label{eq:omegac}
    |V_x|,|V_y| \ll \omega_c = eB/m
\end{equation}
in which we may project in a single Landau level. After projection, the effective Hamiltonian in the $n$'th Landau level is given by~\eqref{eq:AAH1} with $\tilde{V}_x,\tilde{V}_y$ determined by~\eqref{eq:tildeV}. 

We assume that directional localization (and the consequent conductivity anisotropy) is stable to interactions at all, it is most stable at the maximally anisotropic points, where one of $\tilde{V}_x,\tilde{V}_y$ become equal to zero. Consequently we focus our analysis on the vicinity of the point $\tilde V_y = 0$, and later discuss the effect of finite $\tilde V_x$. Due to the $y$ translation symmetry, we adopt a Landau gauge $\vec A = Bx\hat y$ and label states using $y$-momentum $k_y$ and LL index $n$. Let $\varphi_{n,k_y}(\vec r)$ be the Landau gauge orbital with $\tilde x = k_y\ell^2$, namely 
\begin{equation}
    \varphi_{n,k_y}(\vec r) = \frac{C_n}{\sqrt{L_y\ell}} e^{-ik_yy}H_n((x-k_y\ell^2)/\ell)e^{-(x-k_y\ell^2)^2/2\ell^2}
\end{equation}
where $C_n = \frac{1}{\sqrt{\sqrt{\pi}2^nn!}}$. We assign these orbitals creation and annihilation operators with canonical anticommutation relations $\{ c_{n,k_y},c_{n',k_y'}\} = 0,\{ c_{n,k_y}^\dagger,c_{n',k_y'}\} = 2\pi\delta_{nn'}\delta(k_y-k_y')$. Henceforth we leave the dependence on LL index $n$ implicit and drop the $y$ subscripts on all momenta. We would like to study the Hamiltonian 
\begin{equation}
    \hat H = \sum_{k} E(k) c_{k}^\dagger c_{k} + \frac{1}{2} \sum_{q} U(q) :\rho_{q} \rho_{-q}:
\end{equation}
where $E(k)$ is the dispersion and $U(q)$ is the projection of a screened Coulomb interaction $U(\vec r) = \frac{e^2}{\epsilon r} e^{- r/\lambda}$ into the $n$th LL and $\rho_{q} = \sum_{k} c_{k -q}^\dagger c_{k}$. We assume that the typical scale of Coulomb interactions $U_C$ is weak compared to the Landau level bandwidth,
\begin{equation}
    U_C \ll \tilde V_x,
\end{equation}
so that the low-energy physics is dominated by modes within $U_C$ of the Fermi surface. From the form of $E(k)$, we will find that this consists of a set of wires given by the level sets of the dispersion, as indicated in Fig. \ref{fig:contours}. We derive a coupled-wire effective theory from the microscopic model in what follows.
\par 
The microscopic Hamiltonian, after projecting interactions into the $n$'th Landau level and accounting for dispersion, is $\hat H = \hat H_0 + \hat H_\text{int}$ where
\begin{subequations}
\begin{align}
    \hat H_0 &= \int \frac{dk}{2\pi}  E(k) c_{k}^\dagger c_{k}\\
    \hat H_{\text{int}} &= L_y^2\int \frac{d^4k_{i}}{(2\pi)^4} A_{k_{1},\ldots, k_{4}} c_{k_{1}}^\dagger c_{k_{2}}^\dagger c_{ k_{3}}c_{k_{4}}.
\end{align}
\end{subequations}
Here, $A_{k_1,\ldots, k_4}$ is an interaction matrix element whose form we specify later. The low-energy degrees of freedom at generic fillings $\nu$ are chiral edge modes located at some set of positions which we label as
\begin{equation}
    \{\text{wire $x$ positions}\} = \{x_I\} = \{\ell^2 k_{I}\}
\end{equation}
for integer $I$. Specializing now to $V(x) = V_0 \cos(Q_x x)$, denoting the moir\'e period $a_\mathrm{M}=2\pi/Q_x$, and assuming $V_0>0$, the dispersion is 
\begin{equation}
    E(k) = \tilde V_x \cos(Q_x k\ell^2)
\end{equation}
where $\tilde V_x$ is given by Eq. \eqref{eq:tildeV}. The chiral modes at the Fermi surface occur at momenta $k_I$ satisfying
\begin{equation}
   E_F = \tilde V_x \cos(2\pi k_I \ell^2 /a_\mathrm{M})
\end{equation}
which is solved by
\begin{equation}\label{eq:kI}
    k_{I} = \frac{a_\mathrm{M}}{\ell^2} \left(\left\lfloor \frac{I-1}{2}\right\rfloor + \frac12 +(-1)^I\frac{\nu}{2}\right).
\end{equation}
The $y$-velocities are 
\begin{equation}
    v_I = \pdv{E(k)}{k}\left.\right|_{k = k_{I}} \equiv (-1)^I v
\end{equation}
with $v>0$ their magnitude. \footnote{Technically, there are terms generated by normal ordering which we ignore because they are proportional to the interaction scale, which we take to be much smaller than the bare dispersion. Normal ordering generates a 1-body term which dresses the strong superlattice potential $V(x)$ but preserves the translation symmetries. Therefore, it does not alter the starting point for our coupled wires treatment in an appreciable way.} \par 

To proceed with bosonization it is useful to write the Hamiltonian in real space. Therefore we define the fermionic creation operator $\psi(y) = \int \frac{dk}{2\pi} c_{k} e^{ik y}$ satisfies $\{\psi(y),\psi(y')\} = 0,\{\psi(y),\psi(y')^\dagger\} = \delta(y-y')$. Confining ourselves to the vicinities of the wires, we decompose this operator into contributions from each wire \cite{Giamarchi2004}:
\begin{equation}
    \psi(y) \simeq \sum_I \int_{k_{I}-\Lambda}^{k_{I}+\Lambda} \frac{dk}{2\pi} e^{-iky}c_{k} \equiv \sum_I \psi_I(y)
\end{equation}
where $\Lambda \ll a_\mathrm{M}/\ell^2$ is a momentum cutoff. The Hamiltonian restricted to these wires is $\hat H_0 + \hat H_{\text{int}}$ with
\begin{subequations}
    \begin{align}
        \hat H_0 &= -i\sum_I \int dy \, v_I \psi_I^\dagger \partial_y \psi_I\\
       \hat H_{\text{int}} &= \sum_{I,I',J',J} \int dy \, A_{I,I',J',J} \psi_I^\dagger \psi_{I'}^\dagger \psi_{J'} \psi_J
    \end{align}
\end{subequations}
where $A_{I,I',J',J} =  L_yA_{k_{I},k_{I'},k_{J'},k_{J}}$. We define a chiral boson $\phi_I$ for each wire using the mapping 
\begin{equation}
    \psi_I^\dagger = (2\pi \alpha)^{-1/2}e^{-ik_{I} y} e^{(-)^{I+1} i\phi_I} \gamma_I
\end{equation}
where $\alpha \sim 1/\Lambda$ is a short-distance cutoff and the $\gamma_I$ are Klein factors satisfying $\gamma_I \gamma_J = -\gamma_J \gamma_I$. Following a standard procedure \cite{Paul2024Sep}, the bosonized action takes the form 
\begin{subequations}
    \begin{align}
        S &= S_0 + S_{\text{int}} + S_{\text{pert}}, \\
        S_0 + S_{\text{int}} &= \frac{1}{4\pi} \int dtdy\, [\mathsf{K}_{IJ} \partial_t \phi_I \partial_y \phi_J - \mathsf{V}_{IJ} \partial_y \phi_I \partial_y \phi_J]\label{eq:s0sint}
    \end{align}
\end{subequations}
where $\mathsf{K}_{IJ} = (-1)^I\delta_{IJ}$ and $\mathsf{V}_{IJ} = v\delta_{IJ} + (1/\pi)\mathsf{U}_{IJ}$ and $\mathsf{U}_{IJ} =A_{I,J,J,I}-A_{I,J,I,J}$ captures density-density terms. The remaining part of $\hat H_{\text{int}}$, which we package into $S_{\text{pert}}$, is a linear combination of terms of the form
\begin{equation}\label{eq:Sm}
    S_{\vec m} = g_{\vec m}\int dt dy  \,\mO_{\vec m} ,\quad \mO_{\vec m} = \cos(\vec m^T  \vec \phi)
\end{equation}
where $\vec m \in \mathbb{Z}^{2N}$ for a system of $2N$ wires. \par 

In order to consider when the operators in $S_{\text{pert}}$ are relevant perturbations, we must consider their various scaling dimension. The operator $\mO_{\vec m}$ has a scaling dimension given by $\Delta_{\vec m} = \frac12 \vec m^T \mathsf{M} \vec m$
where $\mathsf{M} = \mathsf{A}^T\mathsf{A}$ and $\mathsf{A}$ is a matrix which satisfies $\det \mathsf{A} = 1$, $\mathsf{A K A}^T = \mathsf{K}$ and $\mathsf{A V A}^T = \text{diag}(u_i)$ \cite{Murthy2020Mar}. In order for $\mO_{\vec m}$ to trigger a bulk instability, it is sufficient that two conditions are met: (1) that $\mO_{\vec m}$ is relevant ($\Delta_{\vec m}<2$) and (2) that $\mO_{\vec m}$ is compatible (i.e. commutes) with its translated copies ($\vec m^T \mathsf{K} t^j(\vec m) =0$ for all $j\in \mathbb{Z}$, where $t(\vec m)_I = m_{I-2\text{ mod }2N}$ is a unit cell translation) \cite{Kane2002Jan}.   \par

\par Matrix elements take the form 
\begin{multline}\label{eq:AIJKL}
    A_{IJKL} = \frac{L_y}{2} \int d^2\vec r_1 d^2\vec r_2 \,\varphi_{k_I}^*(\vec r_1)\varphi_{k_J}^*(\vec r_2)\\
   \times U(\vec r_2-\vec r_1)\varphi_{k_K}(\vec r_2)\varphi_{k_L}(\vec r_1).
\end{multline}
An important symmetry of the interactions is $y$ momentum conservation: 
\begin{equation}\label{eq:symmetry}
    k_{I} + k_{J} = k_{K} + k_{L}.
\end{equation}
This, along with charge conservation, places important constraints on the model: namely, perturbations $\mO_{\vec m}$ must satisfy the respective conditions $\sum_I \mathsf{K}_{IJ} m_J k_{J} = 0$ and  $\sum_{I} \mathsf{K}_{IJ} m_J = 0$. Since these are a symmetries of the microscopic Hamiltonian, they must be respected by all terms generated by the RG flow.
\par 
To extend this analysis to study the neighbourhood of the maximally anisotropic points (i.e. to finite $\tilde{V}_y$), we relax Eq. \eqref{eq:symmetry} to a discrete $y$-translation symmetry, to account for scattering off the moire potential
\begin{equation}\label{eq:symmetry}
    k_{I} + k_{J} = k_{K} + k_{L} \mod Q_y.
\end{equation}
In the usual spirit of RG, tuning away from the maximally anisotropic point means turning on small perturbations which break the continuous $y$-translation symmetry down to a discrete $y$-translation symmetry with period $2\pi/Q_y$. Instead of calculating precise effects of such perturbations at the microscopic level, we simply enlarge the space of symmetry-allowed instabilities.

\section{Absence of transverse conductivity}
\label{sec:transverse}
In this section, we analyze whether a transverse conductivity can develop due to interactions, which would signal the breakdown of directional localization. In the absence of interactions, the fixed point theory $S_0+S_{\text{int}}$ has only a longitudinal conductivity (along $\hat y$) at the maximally anisotropic points, and the transverse conductivity is absent. Using general symmetry-based arguments, we now proceed to argue that this also holds in the presence of interactions as well. \par

We first establish this result in the maximally anisotropic limit. Upon Landau level projections, charge conservation and $y$-translation invariance combine into a dipole symmetry along $x$ (i.e. a center of mass conservation) due to $k_I \sim x_I/\ell^2$. This dipole symmetry (which we use interchangeably with $y$-translation symmetry henceforth) forbids net charge transport along $x$. Therefore, a nonzero transverse conductivity requires the breaking of the dipole symmetry. The dipole symmetry, however, remains unbroken even nonperturbatively by an extension of the Hohenberg-Mermin-Wagner (HMW) theorem, applying for dimensions $d\leq 2$ \cite{Kapustin2022Dec}. Therefore, precisely at the maximally anisotropic points, interactions will not induce a transverse conductivity. \par 

We next establish the absence of transverse conductivity holds generically even when the system is perturbed away from the maximally anisotropic points. To see this we consider explicitly breaking the $y$-translation symmetry down to a discrete translation symmetry with period $2\pi/Q_y$. Therefore operators of the form $\psi_I^\dagger \psi_J$ are generated under the condition
\begin{equation}\label{eq:kIminuskJmodQy}
    k_{I}-k_J = 0 \mod Q_y
\end{equation}
and can destabilize the bare fixed point. Using Eq. \eqref{eq:kI}, we see that $k_I - k_J$ generally takes the form $\frac{a_\mathrm{M}}{\ell^2}(n + \eta \nu)$ where $n \in \mathbb{Z}$, $\eta = -1,0,1$, and $\nu \in [0,1]$ is the filling. Let us denote the number of flux quanta per unit cell by $\Phi$, where the unit cell has area $A = (2\pi/ Q_x) (2\pi/ Q_y)$. In general, then, the above is only satisfied for fine tuned conditions, when $\Phi$ and $\nu$ satisfy
\begin{equation}\label{eq:Phinu}
    \Phi = \frac{n+\eta \nu}{m}
\end{equation}
for $m,n\in \mathbb{Z}$. Conversely, if Eq. \eqref{eq:Phinu} is not satisfied,  no coherent transverse tunneling processes are generated by the moir\'e potential and the transverse conductivity remains zero.

Precisely at the conditions Eq. \eqref{eq:Phinu}, the system will flow to a 2D Fermi liquid, inducing a transverse conductivity. We note that even when transverse conductivity is restored, it may be very weak. It arises due to order $m$ Bragg scattering processes and hence for a sufficiently weak (i.e. perturbative) potential $\tilde{V}_y$, we expect the transverse hopping to exponentially suppressed in $m$.

However, most importantly, Eq.~\eqref{eq:Phinu} is generically \textit{not} satisfied, and the system retains its 1D Luttinger liquid qualities, and does not melt into a 2D Fermi liquid. The magnetic field plays an essential role for this, and without it the generic expectation is the opposite \cite{Strong1994Aug,Sondhi2001Jan}. Beyond coupled wire models, the suppression of coherent tunneling in a magnetic field is also generally expected in highly anisotropic materials such as quasi-1D conductors \cite{Lebed2015Oct,Osada1992Jul}. This suppression is almost always promoted to a complete absence in our model, in which the magnetic field is the largest scale (Eq. \eqref{eq:omegac}) and tunneling generically violates the dipole symmetry. 
\par

\section{Survival of longitudinal conductivity}
\label{sec:parallel}

\begin{figure}
   \centering \includegraphics[width=\linewidth]{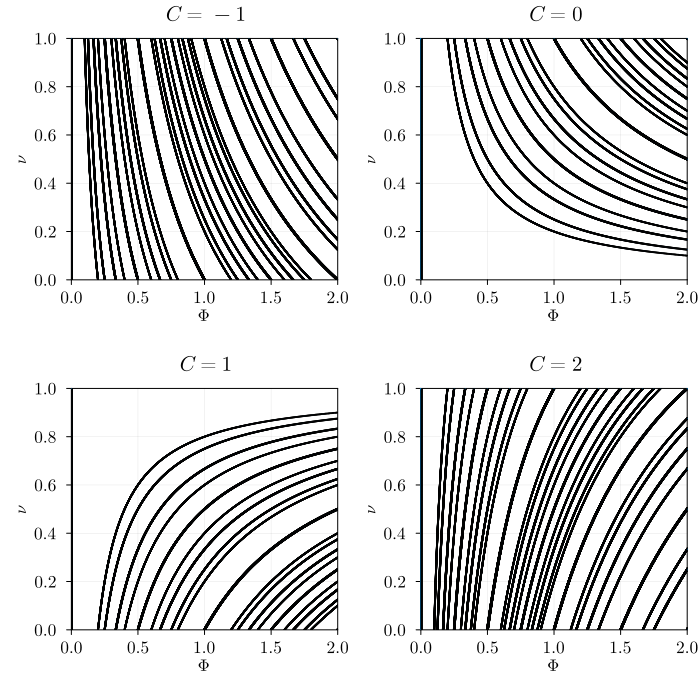}
   \caption{Solutions to Eq. \eqref{eq:phinu2} for $n\leq 5$ and $\nu_{\vec m} = -1,\ldots 2$.} 
   \label{fig:2}
\end{figure}
In this section we address the longitudinal conductivity upon including the interactions. As before, we assess the situation at the maximally anisotropic points first. At these points, the system is an interacting smectic metal with Luttinger liquid-like longitudinal conductivity and zero transverse conductivity. Upon perturbing to finite moire potential $\tilde{V}_y$, we will find that generically, incommensurate CDWs form along $\hat y$ which, in the absence of disorder, allows a sliding longitudinal conductivity through its ungapped Goldstone mode. We demonstrate the formation of these CDWs from microscopic calculations in the next section. Here, we focus on their phenomenology.\par 
We consider first the maximally anisotropic point. At this point the system spontaneously forms a charge density wave, however the conductivity remains finite due to the ``sliding'' Goldstone mode. To see this we note that, as the interactions are weak, it suffices to consider a four-fermion instability. This follows as fermion bi-linear instabilities do not conserve $k_y$, and hence are not generated in the RG by symmetry, while higher-fermion instabilities are always irrelevant at weak interaction strengths. A general symmetry-allowed (i.e. dipole conserving) four-fermion operator takes the form
\begin{equation}\label{eq:I0j1j2}
\mO_{\vec m} \sim \psi_{I_0} \psi_{I_0+j_1}^\dagger \psi_{I_0 + j_1 + j_2}^\dagger \psi_{I_0 + 2j_1+j_2} + \text{H.c.},
\end{equation}
which is understood as before bosonization and up to unimportant prefactors. Here, $j_1,j_2 \in \mathbb{Z}_{>0}$ and $j_2$ is odd. Following Ref. \cite{Paul2024Sep}, when this operator is the most relevant perturbation, the system realises a CDW along $y$ with wavevector 
\begin{equation}
k_{\vec{m}} = \begin{cases} k_{I_0+j_1}-k_{I_0} & j_1 \text{ odd}\\
k_{I_0 + j_1 + j_2} -k_{I_0 + j_1} & j_1 \text{ even}
\end{cases},
\label{eq:cdw_kvectors}
\end{equation}
(i.e. a wavevector obtained by the differences in the $y$ momenta of two fermions occuring in Eq. \eqref{eq:I0j1j2} with opposite chiralities). Eq.~\eqref{eq:cdw_kvectors} may be re-written as
\begin{equation}\label{eq:km}
    k_{\vec m} = \frac{a_\mathrm{M}}{\ell^2} |\nu_{\vec m} - \nu|
\end{equation}
where $\nu_{\vec m}$ is an integer characterizing the phase (the Chern number) and $\nu$ is the filling. Generally, the CDW adjusts its wavevector to satisfy $k_{\vec m} r_c \sim O(1)$ where $r_c\sim \sqrt{n}\ell$ is the cyclotron radius~\cite{Paul2024Sep}. The resulting state has a charge density modultaion $\delta \rho = \rho(y) - \bar{\rho}$ given by
\begin{equation}\label{eq:rhorho}
    \langle \delta \rho(y)\rangle  \sim \cos(k_{\vec m} y)
\end{equation}
where we have assumed an infinitesimal pinning of the CDW. In the absence of pinning, i.e with a \textit{strict} $y$ translation symmetry, there is a Goldstone mode $\phi$ mode which enters as $\cos(k_{\vec m} y + \phi)$, and this ``sliding" mode carries a current in the presence of an electric field \cite{Herbert1954May}. 
\par 
A realistic system will lack strict $y$ translation symmetry, either by deviation from the maximally anisotropic point or an impurity potential, and can cause the CDW to pin and hence the suppression of the longitudinal conductivity. We find that the moire potential is generically unable to pin the CDW due to an incommensuration effect. To see this, let us first consider turning on the modulation $\tilde V_y \cos(Q_y y)$. In terms of the flux per moir\'e unit cell $\Phi = 2\pi /Q_xQ_y \ell^2$, Eq. \eqref{eq:rhorho} reads 
\begin{equation}\label{eq:rhorho2}
    \langle \delta \rho(y)\rangle \sim \cos(\Phi |\nu_{\vec m}-\nu| Q_y y)
\end{equation}
which is generically incommensurate with the underlying modulation $\cos(Q_y y)$ due to the factor $\Phi|\nu_{\vec m} - \nu|$. \par 
\begin{figure*}
    \centering \includegraphics[width=\linewidth]{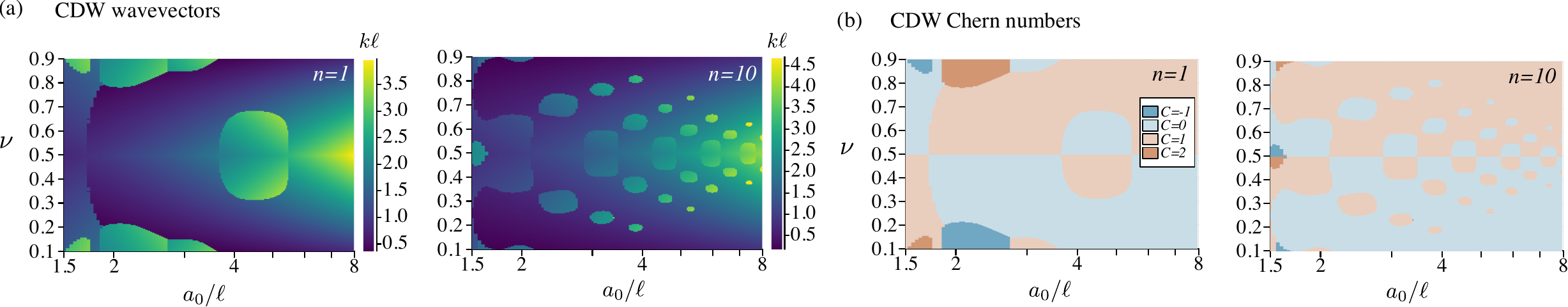}
 \caption{\textbf{Phase diagrams.}  Interacting phases of directionally localized electrons in a magnetic field showing formation of CDWs along the extended direction in the $n=1$ (left) and $n=10$ (right) Landau levels. (a) Spontaneous wavevector of CDW $k$. Note that $k$ varies continuously across much of the phase diagram, so the CDW is generically incommensurate with underlying modulations. (b) Chern number of global state. (Computed with $\ell = 1, d=15, \lambda = 0.5\ell$, $\text{max}[\mathsf{U}_{IJ}/v] = 0.05$.) }\label{fig:phases}
\end{figure*}
\begin{figure}
    \centering
\includegraphics[width=0.7\linewidth]{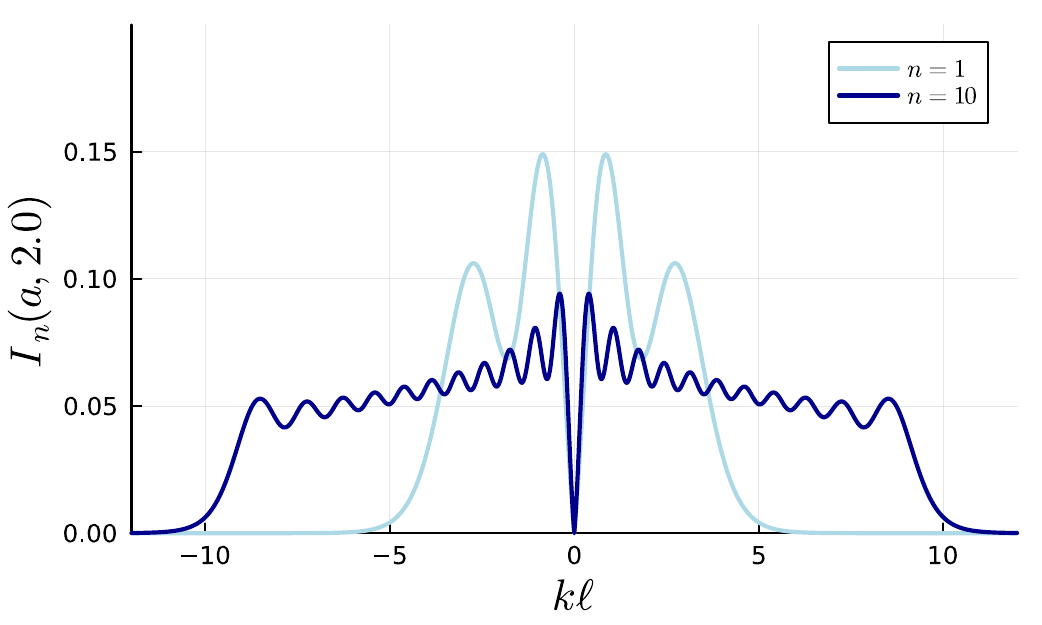}
    \caption{The profile of interwire couplings $\mathsf{U}_{IJ}$ is sampled from the above function for the $n$'th Landau level at screening length $c\equiv \ell/\lambda = 2$ and $k\ell = |k_I-k_J|\ell$. [See Eq. \eqref{eq:Iac} for explicit form of $I_n(a,c)$.]}
    \label{fig:ia2n}
\end{figure}
In a similar spirit to the previous section, there will be a measure zero set of $(\Phi,\nu)$ where commensuration plays an important role. In this case, it is when \begin{equation}\label{eq:phinu2}\Phi |\nu_{\vec m} - \nu| = m/n\end{equation}
for $m,n\in\mathbb{Z}$. Under these conditions, the CDW is the leading instability and the system locks to the underlying modulation, gapping out the Goldstone mode. The set of $(\Phi,\nu)$ for which this occurs depends on the Chern number $\nu_{\vec m}$. We show examples of this set in Fig. \ref{fig:2}. 
\par 

Outside of this set of $(\Phi,\nu)$, the CDW is incommensurate with the underlying moire potential, the CDW remains unpinned, and the longitudinal conductivity remains. A phenomenological picture of this effect establishes that current is carried via the motion of phase-slip ``domain walls'' separating domains of commensurate CDW~\cite{Turkevich1982Apr,Okwamoto1979May,McMillan1976Aug}. The overall state retains a gapless translational Goldstone mode, and hence a finite conductivity.

Finally we discuss the effect of disorder. The effect of disorder on CDWs is extensively studied~\cite{Gruner2019May,Fisher1998Jul}. For sufficiently weak disorder, the dominant effect is causing the incommensurate CDW to pin. In this case, the longitudinal conductivity is restored only above a threshold depinning bias voltage~\footnote{Here we have assumed the disorder is weak enough that the CDW can be treated as elastically deformable classical medium \cite{Fisher1998Jul}}. 

\par 

\section{Incommensurate CDWs}
\label{sec:phases}

We now numerically verify that the ground state of the system is a CDW which is generically incommensurate along the $y$ direction. To this end, we consider the phase diagram of the system as the various microscopic parameters are varied. As $a_0/\ell$ and the Landau level index $n$ change,  the wavefunction profiles and overlaps change. Moreover, adjusting the filling $\nu$ changes the locations of the wires. Altogether, varying these parameters changes the inter-wire interactions $\mathsf{U}_{IJ}$, and hence may alter the leading instability. 

\par To map out the phase diagram, we compute the interactions $\mathsf{U}_{IJ}$ (defined below Eq. \eqref{eq:s0sint}) for each set of microscopic parameters and systematically determine the most relevant symmetry-allowed instability $\mO_{\vec m}$. We work on a system of $200$ wires, and we find that the results do not appreciably change for larger $N$. Fig. \ref{fig:phases} depicts the wavevectors and Chern numbers of the leading instability.  

\par

In particular, Fig. \ref{fig:phases} shows the phase diagram of the $n=1$st and $n=10$th Landau levels with short ranged ($\lambda = 0.5\ell$) Coulomb interactions. The CDWs differ both in their Chern number and in their wavevectors. We note the appearance of ``bubbles" of different phases, which can be attributed to the multiple competing local maxima of the interwire interactions. The interwire interactions $\mathsf{U}_{IJ}$ are obtained from the projected Coulomb density-density interaction and take the form $\mathsf{U}_{IJ} = \frac{e^2L_y}{\epsilon}I_n(|k_I-k_J|\ell,\ell/\lambda)$, where $I_n(a,c)$ is a dimensionless function whose precise form we leave to the Appendix (Eq. \eqref{eq:Iac}). We plot $I_n(a,c)$ in Fig. \ref{fig:ia2n}. As $n$ increases, the interaction has more local maxima and the phase diagram becomes correspondingly more intricate. \par 

An interesting consequence of Fig. \ref{fig:phases} is the following: as we tune magnetic field, not only does the longitudinal conductivity easy axis switch (due to the switching phenomenon at the single-particle level), but the Hall conductivity $\sigma_{xy} = Ce^2/h$ may as well. This is evident from the switching of Chern number $C$ in Fig. \ref{fig:phases}b at fixed $\nu$ as $a_0/\ell\sim \sqrt{B}$ is varied. 

Lastly we remark, for a CDW weakly pinned by disorder, it has previously been noted~\cite{Patri2024Aug} that the nonzero Chern number can lead to distinct behavior of $\rho_{xx}$ at the depinning transition, namely the development of a non-zero $\rho_{xx}$ at a threshold value of $V_y$ with a discontinuous derivative. Moreover, $\sigma_{xy}$ will not be quantized above this threshold.

\section{Discussion}

In summary, we have demonstrated that strain-induced directional localization, initially established as a single-particle phenomenon \cite{Paul2024Jun}, persists even in the presence of electron-electron interactions, and leads to a highly anisotropic conductivity. 
Specifically, we showed that this result holds for the regime of Coulomb interaction $\ll$ moir\'e potential $\ll$ cyclotron frequency, though we anticipate it holds outside this regime as well. 
For instance, it is known from semiclassical arguments~\cite{Paul2024Jun} that single-particle directional localization holds at weak fields as well. 
Here, we have established these results via a perturbative RG analysis of a coupled wires low energy effective model.

Though we note that the absence of transverse conductivity holds nonperturbatively in the interaction strength (in the projected Landau level regime) by an extension of the Hohenberg-Mermin-Wagner theorem to dipole symmetries \cite{Kapustin2022Dec}, the bulk of our analysis was based on a coupled wire model and perturbative RG. It would be interesting to extend this to a strong-coupling approach, especially in light of the closely competing CDW phases we have found, which may frustrate each other and lead to a nontrivial RG flow. \par 
While ours was a continuum treatment, directional localization is also known to occur in lattice models \cite{Barelli1999Dec}, for instance the Hofstadter model on a square lattice with $t_x \neq t_y$. Studying the effect of interactions in such lattice models is a compelling direction for future work.
\begin{acknowledgments}
We thank Gal Shavit and Liang Fu for useful discussions and collaboration on related works, and Chaitanya Murthy for useful discussions. N.P. acknowledges the KITP where this work was initiated. This research was supported in part by the Heising-Simons Foundation, the Simons Foundation, and grants no. NSF PHY-2309135 to the Kavli Institute for Theoretical Physics (KITP). 
\end{acknowledgments}

\label{sec:disc}
\bibliography{bib}

\appendix 

\begin{widetext}

\section{Nature of CDW instabilities}

Since our setup has translation symmetry along $y$ direction, our model has $k$ conservation  
\begin{equation}
    k_{I} + k_{I'} = k_{J} + k_{J'}.
\end{equation}
This, along with charge conservation, combine into dipole symmetry (or center-of-mass conservation) for the coupled-wire effective theory: in particular, allowed interactions are correlated hoppings of a pair of opposite-chirality fermions $(I_0, J_0) \rightarrow (I_0-l, J_0 +l) $. Following Ref. \cite{Paul2024Sep}, they can all be written as 
\begin{equation}\label{eq:I0}
\mO_{I_0,J_0,l} = \int dy \, \psi_{J_0+l}^\dagger \psi_{I_0 - l}^\dagger \psi_{I_0}\psi_{J_0} +\text{H.c.}
\end{equation}
where $I_0<J_0$, $J_0 - I_0 \mod 2 = 1$ and $l>0$. Such terms preserve the center-of-mass position along $x$ direction as required by $y$ momentum conservation. In the thermodynamic limit, $\mO_{I_0,J_0,l}$ and its unit-cell-translated copies induce long-range order (LRO) for the CDW order parameter $\psi_{I_*}^\dagger \psi_{J_*}$, where $I_*$ and $J_*$ are an opposite-chirality pair in $\{I_0-l,I_0,J_0,J_0+l\}$ such that $J_* - I_* \mod 2 = 1$. The resulting CDW has wavevector $Q = |k_{I_*} - k_{J_*}|$ along the $y$ direction. In particular, 
\begin{equation}\label{eq:Q}
    Q = \begin{cases}
        |k_{J_0+l}-k_{I_0}| & l \text{ even}\\
        |k_{I_0}-k_{I_0-l}| & l \text{ odd}
    \end{cases}.
\end{equation}
In general, the resulting CDW is topological. One indication is the presence of $N_{\text{edge}}$ gapless edge modes on a system with a boundary along $x$. Equivalently, $N_{\text{edge}}$ is the number of wires left ``untouched" by by the operator Eq. \eqref{eq:I0} and its translated copies. Some careful counting shows that $N_{\text{edge}}$ takes the values shown in Table \ref{table}. 
\begin{table}
\centering
\begin{tabular}{|c|c|c|c|c|}
\hline
$I_0$ & $l$ &  $N_{\text{edge}} = C$ & $Q\ell^2/a_0 = \Phi_0^{-1}$ \\ \hline\hline
even & even &  $\frac{J_0 - I_0 +l +1}{2}$  &  $\frac{J_0 - I_0 +\ell +1}{2} - \nu$ \\ \hline
even & odd&  $-\frac{l-1}{2}$ & $\frac{l-1}{2}+\nu$ \\ \hline
odd &  even &$-\frac{J_0-I_0+l-1}{2}$&  $\frac{J_0-I_0+l-1}{2} +\nu$ \\ \hline
odd & odd &  $\frac{l+1}{2}$ & $\frac{l+1}{2}-\nu$\\ \hline
\end{tabular}
\caption{Properties of CDW operator $\mO_{I_0,J_0,l}$ (Eq. \eqref{eq:I0}) where $C$ is the Chern number (equal to number of gapless edge modes $N_{\text{edge}}$), $Q$ is the wavevector along $y$, $\ell=$ magnetic length, $a_0=$ period along $x$, and $\nu =$ filling, and $\Phi_0= \#$ flux quanta per unit cell.}
\label{table}
\end{table}

\section{Coulomb matrix elements}

For the interaction, we have used Coulomb matrix elements of the form
\begin{equation}
    A_{1234} = \frac12\int d^2\vec r_1 d^2\vec r_2 \, \varphi_{n,k_1}^*(\vec r_1)\varphi_{n,k_2}^*(\vec r_2)U(\vec r_{2}-\vec r_1)\varphi_{n,k_3}(\vec r_2)\varphi_{n,k_4}(\vec r_1)
\end{equation}
(slightly abusing notation with $k_I \sim I \sim 1$) where $\varphi_{n,k}$ is the Landau gauge wavefunction
\begin{equation}
    \varphi_{n,k}(\vec r) = \frac{C_n}{\sqrt{L_y\ell}} e^{-ik y}H_n((x-k\ell^2)/\ell)e^{-(x-k\ell^2)^2/2\ell^2}
\end{equation}
with $C_n = \frac{1}{\sqrt{\sqrt{\pi}2^nn!}}$. We introduce the parameterization 
\begin{equation}
    k_1 = k-q/2,\quad k_2 = k+q/2,\quad k_3 = k-p/2,\quad k_4 = k+ p/2.
\end{equation}
Then we have \cite{Paul2024Sep}
\begin{equation}
    A_{1234} = \frac{e^2}{\epsilon L_y} \tilde I_n(p\ell,q\ell,\ell/\lambda)
\end{equation}
where we've defined a dimensionless integral 
\begin{multline}
\tilde I_n(a,b,c) = C_n^4\int dS\, ds \, H_n((S+s-a)/2) H_n((S-s+a)/2) H_n((S-s-b)/2) H_n((S+s+b)/2) \\
\times K_0\left(\frac{s}{2}\sqrt{(a+b)^2+4c^2}\right)e^{-\frac{(S+s-a)^2 + (S-s+a)^2+(S-s-b)^2+(S+s+b)^2}{8}}.
\end{multline}
The density-density interaction between two wires is $\mathsf{U}_{IJ} = \frac{e^2L_y}{\epsilon} I_n(|k_I-k_J|\ell,\ell/\lambda)$ where
\begin{equation}\label{eq:Iac}
    I_n(a,c) = \tilde I_n(|a|,-|a|,c)-\tilde I_n(|a|,|a|,c).
\end{equation}
We normalize the set of couplings by setting $\text{max}[\mathsf{U}_{IJ}/v]$ to $g_* = 0.05$. The phase diagrams presented do not differ appreciably for $g_*\in(0,0.1]$ and converge well for $N \gtrsim 25$.

\end{widetext}

\end{document}